# Atomic-state diagnostics and optimization in cold-atom experiments

Krystian Sycz*, Adam M. Wojciechowski, Wojciech Gawlik
*M. Smoluchowski Institute of Physics, Jagiellonian University,*
*Prof. Łojasiewicza 11, 30-348 Kraków, Poland*

We report on the creation, observation and optimization of superposition states of cold atoms. In our experiments, rubidium atoms are prepared in a magneto-optical trap and later, after switching off the trapping fields, Faraday rotation of a weak probe beam is used to characterize atomic states prepared by application of appropriate light pulses and external magnetic fields. We discuss the signatures of polarization and alignment of atomic spin states and identify main factors responsible for deterioration of the atomic number and their coherence and present means for their optimization, like relaxation in the dark with the strobed probing. These results may be used for controlled preparation of cold atom samples and *in situ* magnetometry of static and transient fields.

## Introduction

Precision spectroscopic measurements require adequate control of the quantum state of the atomic sample and of the environment in which atoms/molecules are contained. Representative examples include laser cooling and trapping of atoms[1], optical metrology with atom interferometry[2] and atomic clocks[3], and methods for changing speed of light propagating across atomic sample[4].

Many systems possess long-lived ground states with magnetic degeneracy, where the control is achieved by use of appropriate light, magnetic and radio-frequency (RF) fields to create and manipulate specific spin states[5,6]. Such states are conveniently described by polarization and/or alignment, i.e., appropriate elements of the system's density matrix[7,8].

Particular role in the experiments with superposition states is played by cold atoms which offer unique conditions, such as: (i) slowed-down all relaxation mechanisms, like atomic collisions (particularly the spin-exchange one), (ii) absence of velocity-averaging and related Doppler broadening, (iii) ability to address a single hfs transition, (iv) different atomic dynamics and different stationary states (e.g. lack of inflow of depolarized atoms), (v) ability to work with large optical densities (OD ~100 and more) without experiencing too strong decoherence, last-not-least (vi) possibility of exploring quantum-degeneracy regime, where ultra-low temperatures have, in fact, no alternative.

In this work, we describe preparation of rubidium atoms by laser cooling in a magneto-optical trap (MOT) and a set of diagnostic tools and protocols for precision control of magnetic fields, including non-stationary field monitoring that should prove helpful in many cold-atom experiments in the important temperature range 10-100 μK. In particular, they enable cold-atom vector magnetometry and characterization of tensorial, i.e. based on spatial symmetry, characterization of the atomic superpositions.

The created states are characterized by monitoring the polarization plane of light that passes through the medium. The effect of optical rotation in a magnetic field (Faraday effect) results from light detuning and population distribution[9,10,11]. Sufficiently strong light may not only probe, but also modify the atomic states and create their coherence[12,13,14] leading to the observation of the time-dependent, nonlinear magneto-optical rotation or nonlinear Faraday effect (NFE). This process is closely related with the electromagnetically induced transparency[15], electromagnetically induced absorption[16], coherent population trapping[17] and other manifestations of the interference effects in coherently prepared media[18,19].

## Theoretical background

To relate atomic superposition states with experimentally measured observables we start with a standard relation between the complex refraction index and density matrix (see for example Ref [20])

$$n_{\pm} - 1 \propto \mathcal{E}^{-1} \sum_{e,g} \mathcal{R}e\left(d_{eg}^{(\pm)} \rho_{eg}^{(\pm)}\right), \quad (1)$$

*krystian.sycz@gmail.com.

where $\mathcal{E}$ is the light electric field amplitude, $d_{eg}^{(\pm)}$ are the matrix elements of the dipole moment associated with the $\sigma^{\pm}$-polarized light-beam components linking the ground ($g$) and excited ($e$) states, and $\rho_{eg}^{(\pm)}$ are the related density matrix elements. We consider an atomic system with magnetic degeneracy, i.e. with sublevels $m$ and $m'$ in states $g$ and $e$ of angular momenta $F$ and $F'$, respectively, and assume their linear Zeeman splitting (Fig. 1a).

The quantity directly measured in the experiment is the Faraday angle $\theta$, which determines rotation of the polarization plane of a linearly polarized light beam that propagates through the atomic medium immersed in a magnetic field.

$$\theta \propto \mathcal{E}^{-1} \sum_{e,g} \mathcal{R}e\left(\rho_{eg}^{(+)} - \rho_{eg}^{(-)}\right). \qquad (2)$$

The dependence of optical coherence $\rho_{eg}^{(\pm)}$ on laser frequency is responsible for the spectral dependence of the complex refractive index, which in the case when $e$, $g$ are magnetic sublevels is also influenced by the magnetic field. The latter affects also Zeeman coherences $\rho_{ee'}$ and $\rho_{gg'}$. Consequently, in addition to the resonant factor prior to the summation sign in Eq.(2), optical coherences exhibit also Hanle-like resonances of Zeeman coherences around $B = 0$[21]. These resonances have characteristic widths which reflect relaxation rates $\gamma_m(\gamma_{m'})$ of states $m$ ($m'$)[22,23,24].

For a simple system with $F = 1$ and $F' = 0$ interacting with a longitudinal magnetic field along the quantization axis 0z the relevant sublevels of the ground state $g$ are $|F=1, m=\pm 1\rangle$ and for a small Zeeman splitting ($\omega_L \ll \Gamma$, $\Gamma$ being the natural width of the transition) the rotation angle $\theta$ can be expressed as

$$\theta \propto \frac{\omega_L[\rho_{--}+\rho_{++}+2Re(\rho_{-+})]+\delta(\rho_{--}-\rho_{++})}{\delta^2+(\Gamma/2)^2}, \qquad (3)$$

where $\omega_L = g_F \mu_B B$ denotes the Larmor frequency and $\delta$ is the probe-light detuning from the $F = 1 \leftrightarrow F' = 0$ resonance frequency ($g_F$ is the Landé factor and $\mu_B$ the Bohr magneton). In the absence of coherence $\rho_{-+}$, the first term in Eq. (3) is responsible for the *diamagnetic* Faraday rotation. When population imbalance between the $|-\rangle$ and $|+\rangle$ sublevels is created, the second term in Eq. (3) reflects the, so called, *paramagnetic* rotation which occurs even without external magnetic field[25]. The two kinds of rotation exhibit different symmetries as a function of the light frequency and magnetic field: the *diamagnetic* contribution is characterized by symmetric Lorentzian lineshape as a function of $\delta$, while the paramagnetic one has a dispersive shape which cancels when $\delta$=0 (as a function of $\omega_L$, the paramagnetic rotation is an even function, while the diamagnetic rotation is an odd function).

Optical pumping with circularly polarized light results in a spin *polarization* (non-zero magnetization), described by different populations of individual $m$ magnetic sublevels, i.e. diagonal elements of the density matrix. On the other hand, linearly-polarized light may create *alignment* with no net spin polarization but a specific symmetry in populations of the $\pm m$ sublevels and a certain phase relation between them, i.e., the *coherence,* represented by the non-diagonal elements of density matrix. Due to the symmetry, linearly polarized light couples magnetic states with $|\Delta m|$=2 (Fig. 1a) and greater even numbers for high-intensity (multiphoton) processes[26].

Evolution of atomic observables results from competition between the excitation, decay (relaxation and escape), coherent evolution stochastic noise and various instabilities. Its analysis may be simplified by using a short excitation pulse and interrogation by a weak probe. The short pulse creates elements of the density matrix responsible for state populations and coherences. The relaxation of the excited-stated quantities, governed by spontaneous emission, is usually much faster that the relaxation of the ground-state density matrix elements. Consequently, the elements associated with the excited state decay rapidly to zero, while those associated with the ground-state live much longer. After termination of the pump pulse at $t_0$, the elements associated with Zeeman sublevels undergo the free induction decay (FID) and exhibit the following time dependence

$$\rho_{gg'}(t) = \rho_{gg'}^{(0)} e^{-\gamma_{gg'}(t-t_0)} e^{-i\omega_{gg'}t}, \qquad (4)$$

where $\rho_{gg'}^{(0)}$ is the amplitude of the ground-state density matrix element prepared by the pump pulse at $t_0$ and decaying with rate $\gamma_{gg'}$, and $\omega_{gg'} = \left(E_g - E_{g'}\right)/\hbar$ is the energy difference between states $g$ and $g'$ expressed in frequency units, i.e. the precession frequency of the coherence $\rho_{gg'}$:

$$\omega_{gg'} = |m - m'|\omega_L, \qquad (5)$$

where $m$, $m'$ are the magnetic quantum numbers of the relevant atomic state.

In our experiments, we fix the quantization axis along the probe beam (0z) and focus on two situations: (*i*) precession of an atomic alignment (Zeeman coherences $\rho_{mm'}$ with $|\Delta m|$=2 created by a linearly polarized light) around the longitudinal magnetic field, $B_z$, with frequency $2\omega_L$ and (*ii*) precession of an atomic polarization (coherences with $|\Delta m|$=1 created by a circularly polarized light) around the transverse magnetic field $B_\perp \equiv \sqrt{B_x^2 + B_y^2}$ with frequency $\omega_L$. The latter is associated with magnetic mixing of $m$ sublevels differing by $|\Delta m|=1$ by transverse fields. Consequently, time evolution of the ground-state coherences may exhibit two characteristic frequencies $2\omega_L$ and $\omega_L$ corresponding to polarization and alignment driven by $B_z$ and $B_\perp$, respectively.

In the case of cold-atom experiments, the main relaxation mechanisms include atomic-state losses due to escape of atoms from the observation volume caused by ballistic expansion, optical pumping to non-interacting states or pushing atoms out of the trap, trap imperfections, and/or collisions with rest-gases, as well as dephasing due to magnetic field and light-shift (AC Stark effect) inhomogeneities.

## Methods

The experiments were conducted using cold rubidium atoms (both $^{87}$Rb and $^{85}$Rb, consult Table 1 in Suppl. Info. for figure reference) released from a MOT. Experimental setup and timing sequence diagram are shown schematically in Fig. 1b and c, respectively. Cold rubidium samples are periodically released from the trap. After the MOT fields shut off, a specific spin distribution in the ground state is prepared using a short circularly- or linearly-polarized pump pulse, resonant with the trapping transition. Atoms then undergo precession in the applied magnetic field, while their interaction with stray fields is minimized by magnetic shielding of the trap. Independent beam with a variable detuning probes the birefringence of the expanding atomic cloud. The rotation of probe-beam polarization plane is recorded using precision balanced polarimeter and fast photodetectors. Finally, the trap is switched back on and the cycle is repeated.

Measured birefringence provides information on the ground state polarization and coherence and their evolution in the magnetic field. In a single beam experiment, the pumping pulse is omitted, and signals arising solely due to the interaction with a probe beam are recorded. Some experiments were performed with delayed or strobed probing to minimize the effects of light on the state evolution. More detailed description of experimental setup and procedures is enclosed in the Supplemental Information. Datasets acquired in this study are available from the corresponding author on reasonable request.

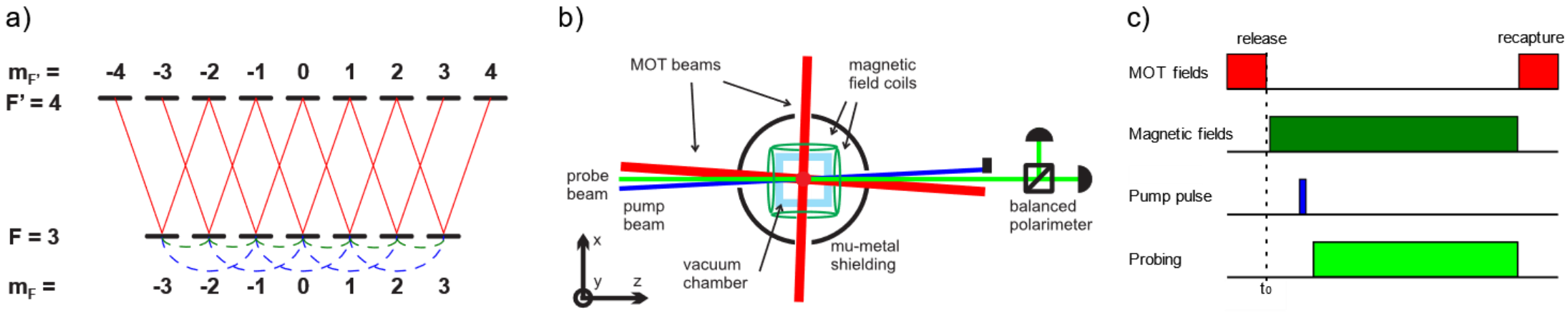

**Fig. 1**. Schematics of the experimental arrangement. (a) A diagram of relevant Zeeman levels of $^{85}$Rb, showing optical transitions caused by the linearly polarized light (red lines) and the $|\Delta m|=2$ coherences it creates (dashed blue). Green dotted lines indicate $|\Delta m|=1$ coherences created by the RF or static perpendicular field. (b) schematic diagram of the experimental setup. (c) timing sequence of the applied fields.

## Results

### Single-beam magneto-optical rotation

With a single linearly polarized beam two types of rotation resonances occurring around the zero-field value can be observed, shown in Fig. 2 as a function of the magnetic field $B_z$. A broad (several Gauss wide) resonance (Fig. 2a) results from a linear Faraday rotation, while the central, much narrower resonance (Fig. 2b) reflects the NFE caused by the Zeeman coherence [12, 13]. Figure 2c shows the observed signals measured for different interaction times of atoms with light. The inset shows the expanded region of $B_z \approx 0$, with the distinct contribution of NFE caused by Zeeman coherence. While the broad (about 6 G wide), linear (with respect to the light intensity) rotation signal appears instantaneously after the measurement begins (blue curve), the narrow NFE resonance takes some build up time that depends on the beam intensity. The NFE resonance shown in Fig. 2b has a FWHM of 6.5 mG, which corresponds to a coherence lifetime of ~100 μs. Figure 2d shows the rotation angle as a function of time for three values of $B_z$. The top curve ($B_z$ = 8 mG) illustrates the buildup of the nonlinear effect on a short time-scale of about 2 ms, which is then limited by the atom-number losses caused by the light (heating and the related loss of coherence), freefall and ballistic expansion. The middle ($B_z$ = 2.2 G) and the lowest ($B_z$ = 5.4 G) curves show the immediate onset of the linear effect and its domination at short times. At longer times (~1 ms) the signals reflect changes of the population distribution caused by optical pumping, seen as inversion of the rotation sign of the middle trace. Eventually, for times $\gtrsim$5 ms the rotation signals tend to zero.

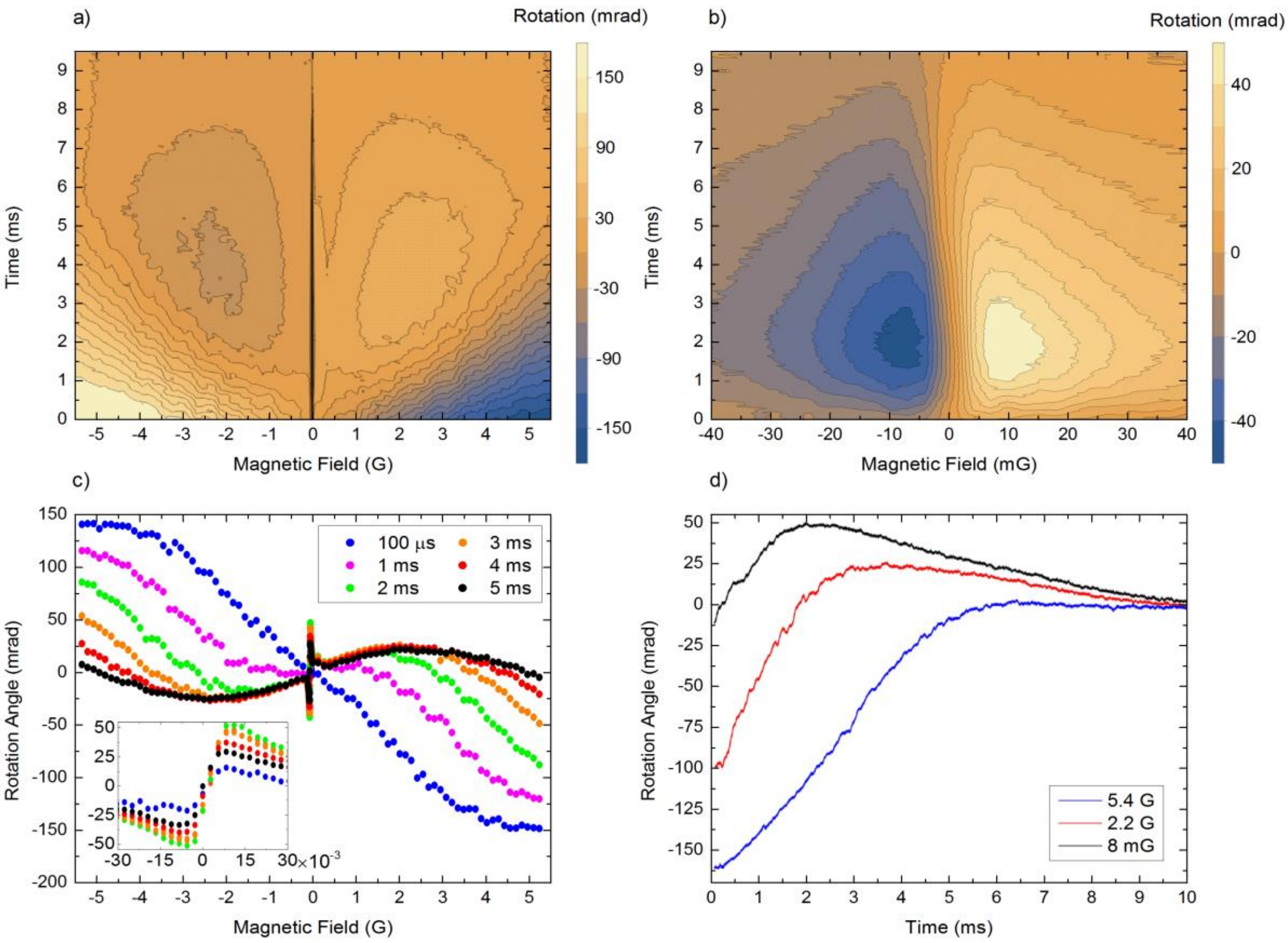

**Fig. 2** Faraday rotation observed with a single beam of 2.5 μW/mm$^2$ intensity, tuned to resonance. a) Contour plot of the rotation angle dependence on magnetic field and time after switching on the light. b) expanded narrow central region of $B_z \approx 0$. c) horizontal cuts from a) showing the rotation signals after 1, 2, 3, 4, 5 and 6 ms (top to bottom traces on the left side, respectively. d) Time evolution of the rotation signal $\theta(B_z)$ for magnetic-field values of: 8 mG (top black), 2 G (middle red), and 5.5 G (lowest blue).

The narrow central part of the rotation curve that corresponds to the NFE rotation (Fig. 2b and inset to Fig. 2c) enables characterization of the Zeeman coherence of trapped atoms [12] and precise measurements of weak magnetic fields[12,14]. In particular, for fields where the Zeeman splitting is smaller than the relaxation rate γ of the atomic coherence, $g_F \mu_B B < \gamma$, the linear dependence $\theta(B_z) \propto B_z$ enables direct measurement of the field intensity if proper calibration is performed. Sensitivity of such measurement improves with increasing amplitude and decreasing width of the NFE resonance. In the case of cold atoms, the signal amplitude and hence the sensitivity changes over time. This differs from the case of hot atomic vapours, where stationary rotation can be observed as a result of atoms flying in and out of laser beam. The maximum rotation amplitude observed in a single beam experiment was of the order of ~200 mrad for intensities above 10 μW/mm$^2$, at the cost of significant power broadening of the resonances.

The width of the NFE resonance corresponds to the effective decoherence rate $\gamma$, which results from various contributions. The most dominant are: light-induced losses $\gamma_{light}$, which depend on the light intensity (scattering rate), the atom-loss rate $\gamma_{TOF}$ due to freefall and ballistic expansion of escaping atoms, and magnetic-field related decoherence $\gamma_B$ due to uncompensated transverse fields, longitudinal field gradient as well as the oscillating fields and the field noise. The recorded resonance width reveals an effective decoherence rate $\gamma = \gamma_{TOF} + \gamma_{light} + \gamma_B$. When some of its constituents differ strongly, it may be possible to identify the dominating decoherence process and minimize it iteratively to optimize the net coherence lifetime.

**Pump-probe measurements**

Compared to the single-beam experiments, the use of separate beams for creation and probing of atomic states offers additional flexibility of pump polarization, detuning and intensity. Most importantly, however, it offers longer lifetimes of atomic polarization and coherence. In two-beam experiments a short pumping pulse creates a specific population distribution in the cold-atom sample and probe light is then used to record the time-dependent rotation. To minimize the probe's influence on atoms, it can be either attenuated, detuned from the optical resonance, and/or used in a strobed manner (see Supplemental Information). Signals associated with atomic polarization could be probed for a wide range of probe-beam detunings (up to several GHz). Already at 70 MHz detuning to the blue, and ~1 μW of light power rotation signals can be continuously probed for 15 ms without a discernible effect on the decay time. On the other hand, detection of atomic coherence is sensitive to the probe frequency, i.e. the rotation amplitude falls rapidly when the detuning increases above the natural width (Eq. (3)).

**FID of atomic polarization**

Optical pumping by circularly polarized pump pulses propagating along the 0z direction, creates a population imbalance between sublevels $m$ and $-m$ that is equivalent to the sample's magnetization, or polarization along the pump beam, $\langle F_z \rangle \neq 0$, described by diagonal elements of density matrix.

Arbitrarily oriented DC magnetic field affects the atomic polarization induced by circularly polarized pump pulses in two ways: ($i$) When $\boldsymbol{B} \nparallel 0z$, the transverse components of the magnetic field mix the adjacent ($|\Delta m|$=1) sublevels. The created atomic polarization undergoes Larmor precession with frequency $\omega_L = g_F \mu_B B$ proportional to the total field strength $B = \sqrt{B_x^2 + B_y^2 + B_z^2}$, with amplitude dependent on the direction of the field[27]. The sign of the rotation depends on the pump beam helicity (left- or right-handed) and on the sign of the probe beam detuning (Fig. 3b). ($ii$) For $B_x = B_y = 0$ and $B_z \neq 0$, there is no oscillation – the magnetization is stationary, and its decay is caused primarily by the decreasing number of atoms.

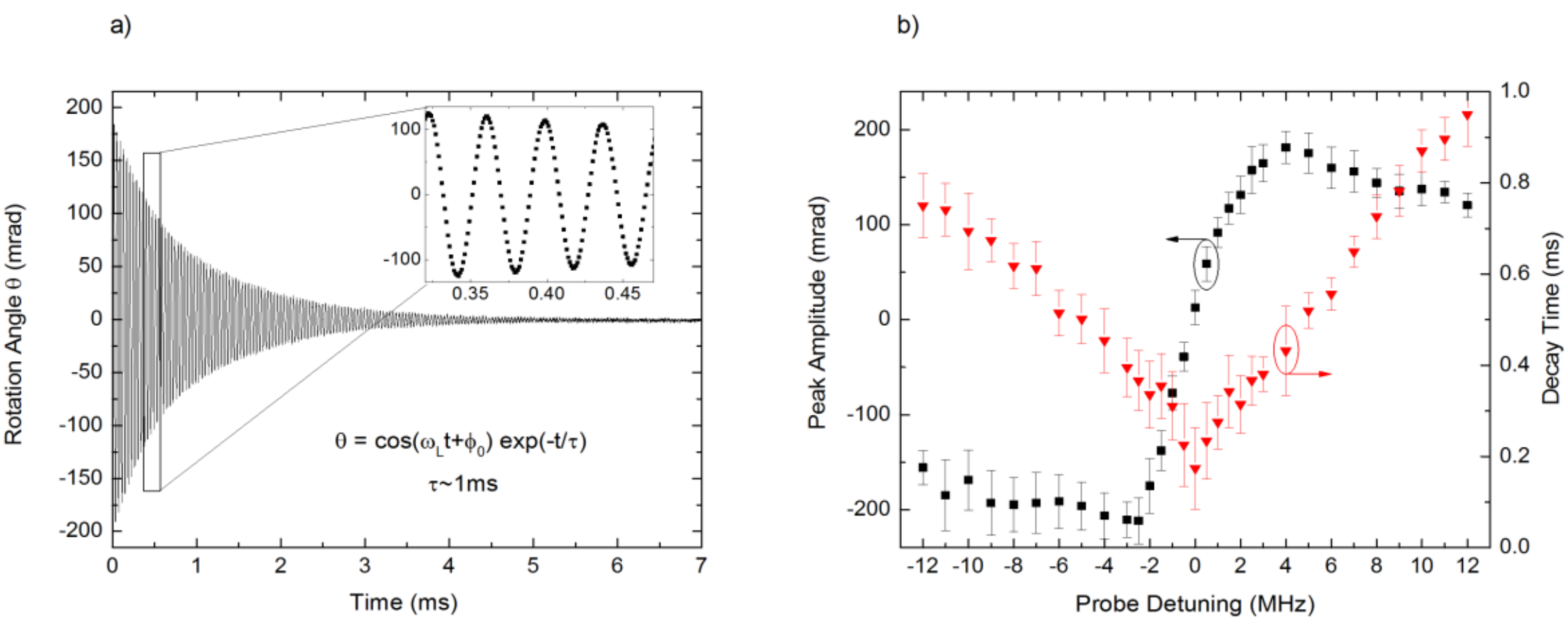

**Fig. 3** FID of atomic polarization. a) Example signal for $B_x$ = 56 mG, with the probe 70 MHz blue detuned from the atomic transition. b) Amplitude (red triangles, left scale) and decay time (black squares and right scale) of the fitted FID signals near the atomic resonance vs the probe detuning $\delta$. Data points are the mean values of 5 non-consecutive measurement series while the error bars show the standard deviation and are indicative of variations in the trapped atom number.

Figure 3a depicts precession of the atomic polarization observed after applying a relatively strong (~1 mW/mm^2) and short (< 20 μs) pump pulse, while switching on the homogenous magnetic field $B_x$ =56 mG, along the 0$x$ direction at $t$ =0. The resulting rotation signal represents FID of atomic polarization, which oscillates at the Larmor frequency $\omega_L$, which is the measure of the intensity of total magnetic field, while its decreasing amplitude reflects polarization relaxation and the decay of the number of interrogated atoms. Figure 3b illustrates the dependence of the amplitudes and decay times of the polarization FID signals on the probe detuning. Typically for paramagnetic rotation, the FID amplitude has a dispersive dependence on $\delta$ with a width close to $\Gamma/2\pi$=6 MHz and vanishes at $\delta$=0. Its decay exhibits strong decoherence by the resonant probe beam.

FID signals can be a useful indicator of the magnetic field alignment, homogeneity, and/or presence of parasitic stray fields. In the ideal case of magnetic field perfectly aligned along 0$x$, the FID signal is symmetric around zero value with exponentially decaying envelope. Stray fields, gradients and instabilities cause non-exponential decay, asymmetry and variations in the oscillation frequency (Suppl. Inf.).

**FID of atomic coherences**

Setting the pump beam polarization to linear enables the creation of Zeeman superpositions. Here, we focus on the coherences of Zeeman sublevels described by non-diagonal elements of density matrix with $|\Delta m|$ = 2. The coherences contribute to atomic alignment and can be monitored by a linearly polarized probe beam as a rotation of its polarization

plane[12,13,14]. In time-dependent measurements atomic alignment manifests as oscillations at the frequency of $2\omega_L$ and are usually damped faster than polarization signals (Fig. 4a, b).

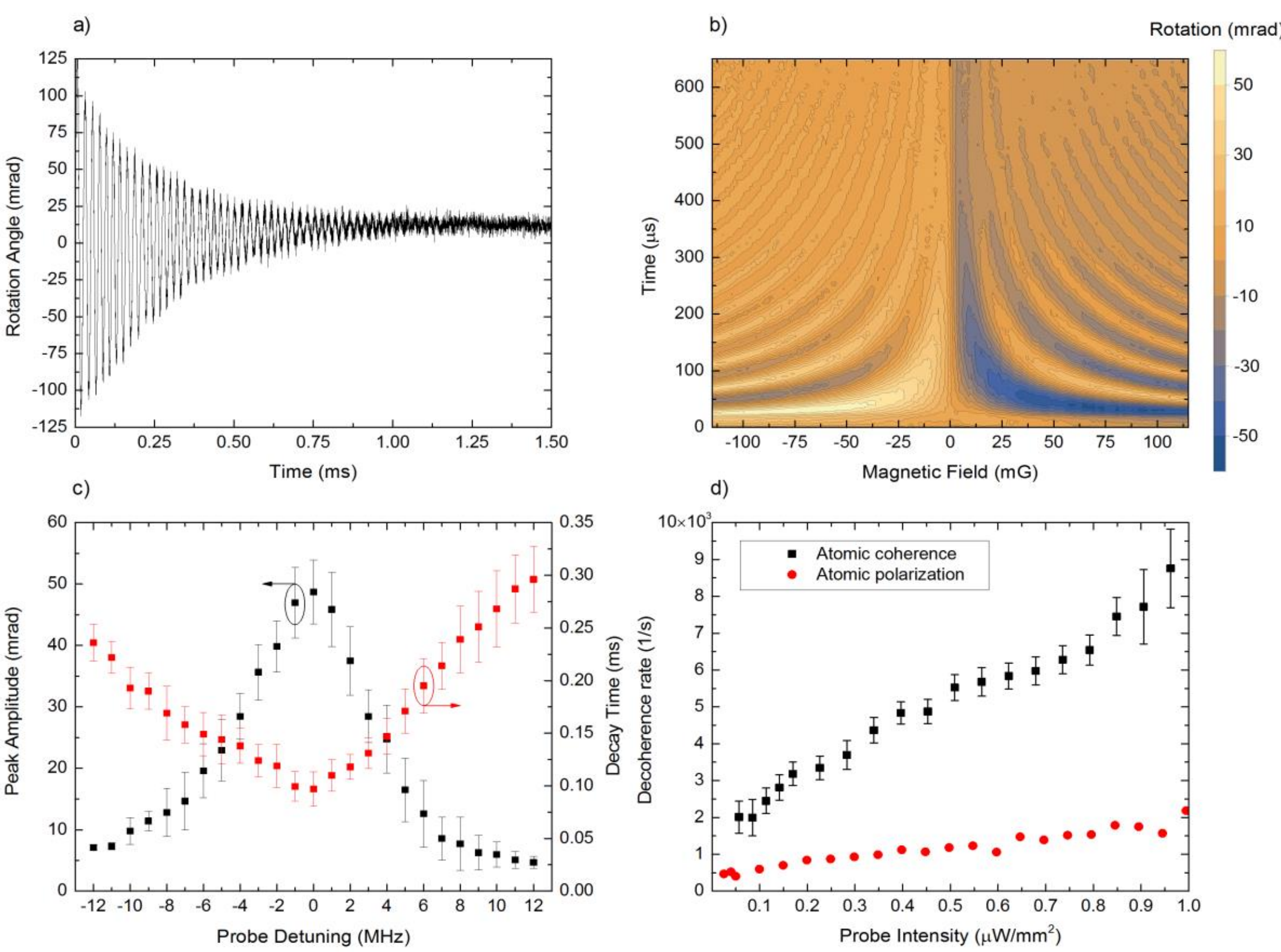

**Fig. 4** FID signal of atomic $|\Delta m|$=2 coherence created by a linearly-polarized pump pulse propagating along the probe beam: (a) time dependence of the rotation signal $\theta(t)$ in magnetic field $B_z$ = 95 mG, under continuous probing. (b) FID signal of atomic coherence $\theta(B_z, t)$ as a function of time and $B_z$. (c) the amplitude and decay time of the coherence FID signals dependence on the probe detuning. (d) decay rates of atomic polarization and coherence on the probe light intensity (CW probing with light detuned by 24 MHz to the blue). Error bars like in Fig. 3.

While the creation and observation of Zeeman coherences under CW conditions is relatively straightforward and can be easily monitored as narrow NFE signals around $B_z \approx 0$ (Fig. 2), observations of these coherences at $B_z \neq 0$, and particularly their FID signatures, is by far more demanding. Transverse magnetic fields have to be carefully compensated and possibly the whole system shielded from any AC fields. The amplitude of the coherence signal rapidly decreases with the probe-beam detuning $\delta$ (black squares in Fig. 4c), thus, $\delta$ should be small, on the order of Γ. Unfortunately, small $\delta$ enhances decoherence of the alignment (red triangles in Fig. 4c). It is therefore optimal to pump with the resonant beam, and trade-off the intensity and detuning of the probe. Figure 4c depicts Lorentzian symmetry of the observed coherence FID signals, characteristic for diamagnetic rotation. It is usually beneficial to adjust the pump-pulse duration to the value of Larmor frequency. We observed that the decay rates of the atomic polarization and coherence differ strongly as shown in Fig. 4d for both signals recorded under identical probing conditions. The coherence decay time is consistently faster than for the polarization FID, and both increase at different rates with probe intensity. The slowest observable decay of coherence corresponded to approximately 0.5 ms. Observation of longer coherence lifetimes requires further probe beam attenuation, and may become impractical due to degradation of the signal-to-noise ratio. For this reason, we have employed the stroboscopic technique of relaxation in the dark discussed in Supplementary Information.

### Zeeman superpositions in tilted magnetic field

When the magnetic field is tilted with respect to the probe field direction 0*z* the evolution of the Zeeman superpositions becomes more complicated. As described in Ref. [28] , when the magnetic field vector is tilted away from the 0*z* direction, in the plane of pump light polarization and propagation direction (YZ, angle α), the rotation becomes a superposition of two damped sine functions, one oscillating at $2\omega_L$ and the other at $\omega_L$. Figure 5a depicts the rotation signal recorded in such a tilted magnetic field. In addition to the $|\Delta m|$=2 coherence signal that oscillates at $2\omega_L$, a component oscillating at $\omega_L$ is also visible and can be interpreted as a manifestation of superpositions with $|\Delta m|=1$. Since both components originate from the same atomic population distribution, they decay at the same rate. The amplitude of the $2\omega_L$ contribution diminishes with the tilt angle $\alpha$, while the other one increases. From $\alpha = 45$ deg the contribution at $\omega_L$ becomes dominant. The pump-probe setup allows us to observe the described evolution in real time, and the analysis of such signals can be used to determine the tilt angle (black squares in Fig. 5b). Tilting the field in the direction perpendicular to both, the pump light polarization and propagation direction (XZ, angle $\beta$), does not introduce the modulation at $\omega_L$, but reduces the oscillation amplitude (red triangles in Fig. 5b). With appropriate calibration, it is possible to exploit this effect for vector magnetometry. Similar results have been discussed by the Weis and Riis groups[29,30], while an alternative method of vector magnetometry has been proposed in Refs. [31, 32]

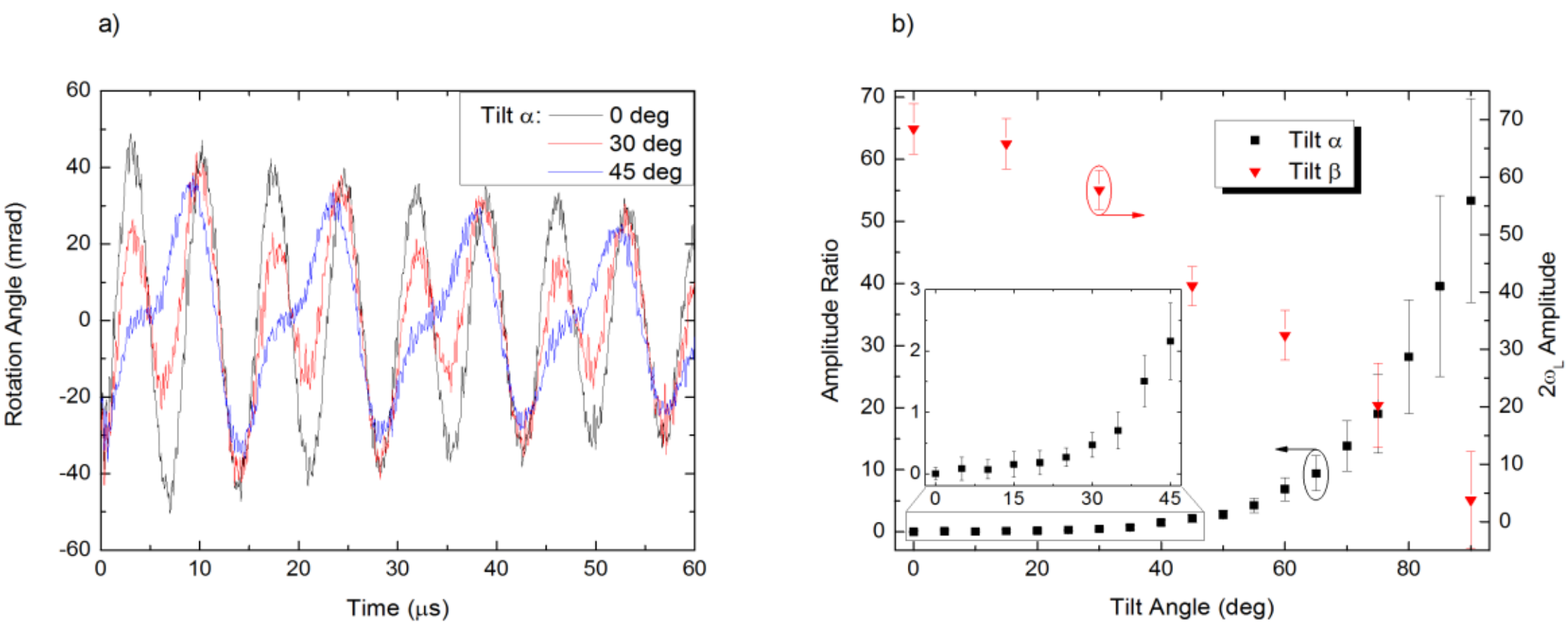

**Fig. 5**. Coherence FID signals in tilted fields (with pump light polarized in the $0y$ direction). a) Modification of the signals as the field is tilted away from the $0z$ direction and towards $0y$ by an angle $\alpha$. A bimodal character is clearly visible. b) The ratio of the amplitudes of components $A(\omega_L)/A(2\omega_L)$ changes as $\tan\alpha$ and can be used to determine the tilt angle (black squares). The inset shows the zoomed in low tilt angle region. Tilting of the field towards the $0x$ direction results only in diminishing amplitude of the oscillations $A(2\omega_L)$ as the $\cos\beta$ (red triangles). Error bars like in Fig. 3.

### Detection of radio-frequency fields

An unequal distribution of atomic population among Zeeman sublevels caused by optical pumping enables us to use the magnetic resonance technique, known as radio-optical double resonance or optically detected magnetic resonance (ODMR). ODMR in a system of Zeeman sublevels is a popular method applied in many situations including cold-atom magnetometry[14]. Below, we present an application of the technique for detection of RF magnetic fields. A short circularly polarized pump pulse creates an atomic polarization along the probe beam ($0z$ axis, as in the case of atomic polarization), while magnetic field $B_z$ splits the sublevels by $\omega_L$. A transverse RF field of frequency $\omega_{RF}$ matching the Zeeman splitting alters the population distribution which can be monitored via rotation of the probe beam polarization.

Depending on the Rabi frequency for the RF field-atom interaction, $\Omega_R = \mu_B B_{RF}$, where $B_{RF}$ is the amplitude of the magnetic component of the RF field, two different regimes of the RF field interaction with atomic sample can be investigated: the weak- and the strong-field one. Figure 6b presents contour plots illustrating the ODMR rotation resonances for both regimes with RF frequency on the order of tens of kHz as a function of $B_z$ and time. In the weak-field regime, where $\Omega_R < \gamma_{gg'}$, the ODMR signals manifest as vertical stripes occurring at two resonant frequencies, $\omega_{RF} = \pm\omega_L$. By tuning $\omega_{RF}$ to the Zeeman splitting of $\pm\omega_L$ one can measure the magnetic field intensity.

In the strong field regime, $\Omega_R > \gamma_{gg'}$, the stripes as a function of time exhibit distinct structure of Rabi oscillations (Fig. 6b) with frequency

$$\Omega'_R = \sqrt{(\omega_{RF} - \omega_L)^2 + {\Omega_R}^2}. \qquad (6)$$

Figure 6b reveals also the characteristic bell-shape bending of the oscillation fringes around the resonance centre which is due to the difference between the resonant and non-resonant values $\Omega_R$ and $\Omega'_R$. By measuring the Rabi frequency, we are able to measure the RF field intensity. Assuming that we need at least one period of Rabi oscillations to determine $\Omega_R$, the decay time of the polarization sets the floor of 150 μG on the measurable RF intensities in our current setup.

Like in the case of coherence FID signals damping rate of the observed Rabi oscillations is also affected by the probe-beam intensity. The relaxation in the dark method can be used to maximally extend the available observation time, as seen in Fig. 6c.

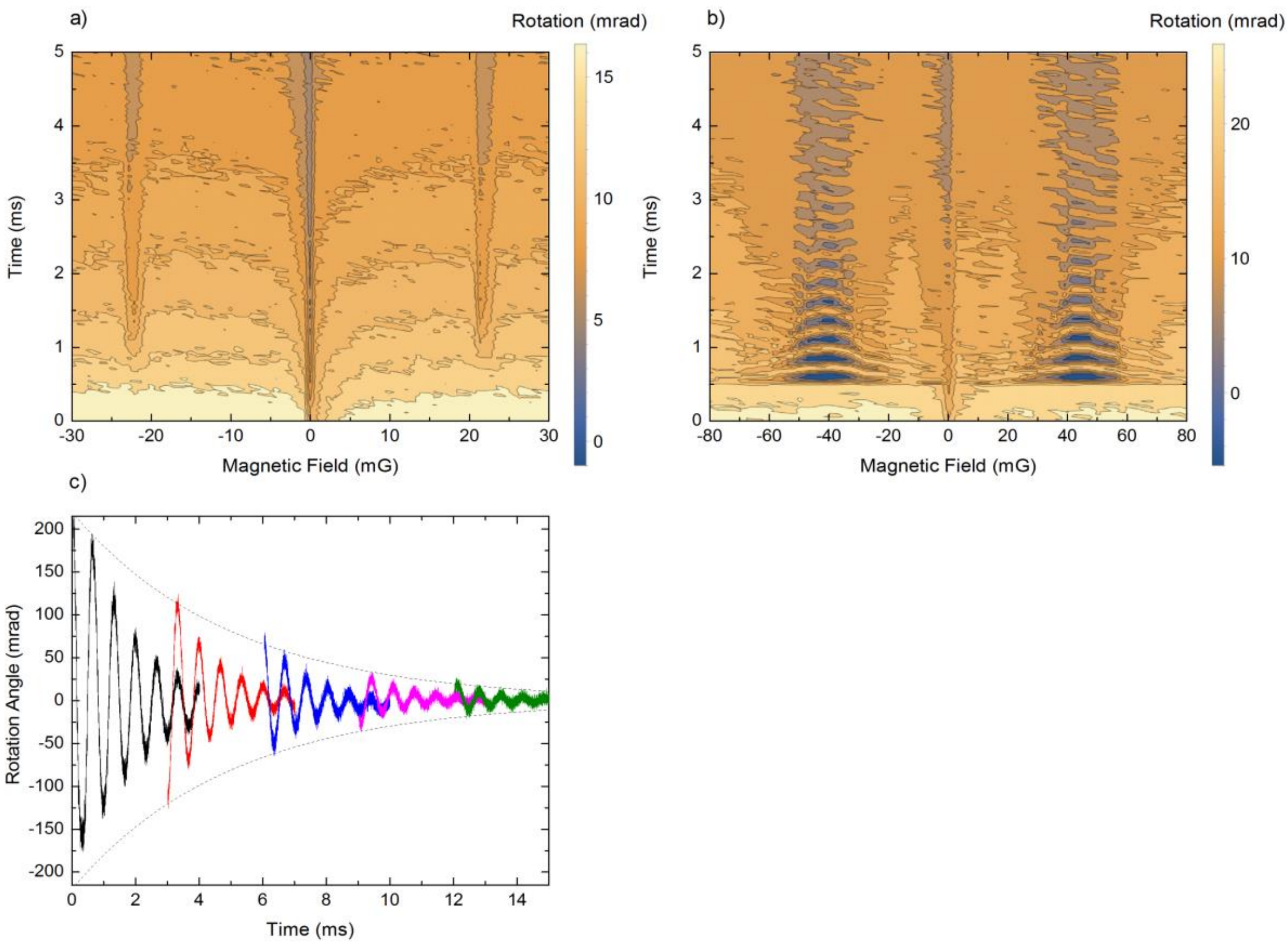

**Fig. 6**. RF induced ODMR signals recorded as a function of $B_z$ and time in two RF-intensity regimes: $\Omega_R \lesseqgtr \gamma_{gg'}$. (a) Weak-field regime, $B_{RF} \approx$ 200 μG at $\omega_{RF}$=2$\pi$ 10 kHz. The dashed line indicates time at which RF field is turned on. (b) Strong-field regime, $B_{RF} \approx$ 7 mG at $\omega_{RF}$=2$\pi$ 20 kHz, revealing RF induced Rabi oscillations seen as horizontal bell-shaped fringes. Minimum oscillation frequency is around +/- 43 mG. The central feature in both figures is an artefact of the sample preparation process. c) A signal of RF induced Rabi oscillations observed by the relaxation in the dark method. The individual signals decay with $\tau$ = 1.56 ms, while the envelope with $\tau$ = 5 ms; a rate that is comparable with the calculated rate of atoms leaving the interaction area.

# Discussion

We have presented a set of simple yet versatile diagnostics and optimization tools for experiments with cold atoms where well-controlled atomic superpositions could be produced and characterized. The fundamental limitation of the coherence lifetime of atoms released from a trap is their ballistic expansion and escape from the detection region. This may be substantially extended by using conservative traps which enable interaction with trapped atoms and offer trapping times on the order of minutes[33]. The described methods could then contribute to achieving very long coherence lifetime and, hence, much higher magnetometric sensitivity.

We have demonstrated that even on a 10 ms time scale of ballistic expansion from the MOT several decoherence processes which affect superposition states can be identified. For superposition of states with differing magnetic quantum numbers important decoherence mechanisms are the inhomogeneity and instability of the external magnetic field (see also Figs. S2 and S4 in Suppl. Inform). These may be suppressed by a proper magnetic shielding and usage of stable current sources for field generation. Another major source of decoherence is the effect of the probe beam intensity, that can be reduced by detuning the light from resonance, at the expense of the amplitude of coherence signals. An alternative and practical solution is the strobing of the probe light to minimize the interaction time (see also S3 of Suppl. Info.) while maintaining low detuning and relatively high intensity.

The described experiments demonstrate that the coherence/superposition states of investigated cold-atom samples live shorter than the states characterized by atomic polarization (populations). Sensitivity of the former ones to light and magnetic field stability and uniformity requires the usage of magnetic shielding, mains synchronization, observations with strobe light or the relaxation in the dark method in order to observe a comparable lifetime.

We have also demonstrated *in situ* diagnostics of the magnetic fields in the cold atom sample region, using magneto- optical rotation caused by atomic polarization and superposition states. These methods can be used for a wide range of cold-atom experiments, like for precise monitoring of atomic states during quantum state manipulation experiments, calibration of magnetic field sources, compensation of external fields, or for laser detuning diagnostics.

**Acknowledgements**

This work was supported by the Polish National Science Centre (grants 2012/07/B/ST2/00251 and 2016/21/B/ST7/01430) and partly by the European Regional Development Fund in the framework of the Polish Innovation Economy Operational Program (contract POIG.02.01.00-12-023/08).

**Author contribution statement**

A.M.W and W.G. conceived the study. K.S. performed experiments, and all authors analysed data, discussed the results and wrote the manuscript.

**Additional information**

The authors declare no competing financial interests.

## Supplemental information

**Experimental arrangement**

The experiments were conducted using cold rubidium atoms released from a magneto-optical trap. Two-cell-system prepared $10^8$ to $10^9$ cold atoms of either $^{85}$Rb or $^{87}$Rb (for reference see Table 1) in a volume of a few $mm^3$ at about 100 μK in ultra-high vacuum (initial vacuum pressure below $10^{-10}$ mbar).

The cooling and trapping light for both the 2D and 3D MOT, as well as for the pushing beam, was delivered from a commercial tapered amplifier laser system. A home-built diode laser provided the repumping light. An independent external-cavity diode laser was used for measurements of magneto-optical rotation and atomic state preparation. Its frequency was stabilized with either a saturated absorption or a Doppler-Free Dichroic Laser Lock (DFDL)[34] spectroscopy setup. Light from this laser is split into two beams, the pump and the probe, both about 1.5 mm in diameter at the atomic cloud volume. Their intensities and detunings are independently controlled by acousto-optical modulators (AOMs), each being in a double-pass configuration. Both laser beams can therefore be precisely tuned around the desired transition of either of the rubidium isotopes. Results presented in this work refer to the rubidium $D_2$ line cyclic transitions: F'=4 – F=3 for $^{85}$Rb (Figures: 1-3a, 4a-b, 6, S1-S5) and F'=3 – F=2 for $^{87}$Rb (Figures: 3b, 4c-d, 5).

Care was taken to precisely compensate any transverse fields and possibly shield the system from stray fields (AC and DC). The science chamber is isolated from external magnetic fields by a cylindrical, single-layer μ-metal shielding of 10 cm in both length and diameter, and a thickness of 0.5 mm. Optical access for laser beams and imaging optics, as well as opening for the vacuum apparatus is provided by means of protruding μ-metal sleeves, about 3 times longer than the diameter of the respective opening. The shield provides a DC shielding factor of about 20. Inside the shielding, well-controlled magnetic fields are generated by six pairs of coils which create uniform and gradient magnetic fields. Home-built current sources control the magnetic field with precision better than 1 μG and perform switching of both AC and DC fields or their superposition within about 0.1 ms. To increase the long-term current stability, currents are not switched off but directed to sink circuits.

A high-quality crystal polarizer linearly polarizes the probe beam just before it enters the science cell. After passing the sample, beam polarization is analysed with a balanced polarimeter, consisting of a Wollaston prism and two avalanche photodiodes operating in a linear intensity-response mode. Simultaneous recording of both photocurrents allows us to measure the optical-rotation angle. Alternatively, the forward scattering geometry may be used with only one detection channel with a polarizer orthogonal to the linear polarizer[35]. The pump beam is crossing the sample either at a small (<2 deg) or at right angle with respect to the probe beam (along $0x$). Its polarization is set to either linear or circular. Both beams propagate at a small angle with respect to the normal of the science cell walls in order to avoid interference effects from the reflected beams (Fig. 1b of the *main article*).

### Single-beam and pump-probe configuration

Two classes of measurements were performed: one with a single beam that simultaneously created and probed given atomic states and second, the pump-probe measurement, where a weak probe interrogated the state prepared earlier by a pump beam. Depending on the polarization of the pump beam and direction of the magnetic field, in the two-beam, pump-probe measurement we observed the time-dependent free induction decay signals associated with either atomic polarization or alignment (coherences with $|\Delta m|$=2).

### Measurement sequence

Timing sequence of the experiment (Fig. 1c of the *main article*) starts with loading of the main trap from the 2D MOT and through recapturing the expanding atom cloud from a previous measurement. After the loading time, the trapping optical and quadrupole magnetic fields are turned off while the desired homogenous magnetic fields are turned on. From this moment, the atomic cloud starts to expand and fall freely, while the actual rotation measurements are delayed by 5 ms to allow for the eddy currents (induced mostly by the switching of the MOT quadrupole field) to decay and the applied fields to stabilize.

In the single-beam experiments, one linearly-polarized, near-resonant beam parallel to the magnetic field oriented along the z-axis performs the readout and manipulation of the atomic state. After switching on the beam, it's polarization rotation is recorded by the polarimeter. Such experiments are referred to as measurements of the NFE.

In two-beam experiments, atoms are initially prepared in the desired state by a relatively short (<15 μs) and intense (~1 mW) pumping pulse of appropriate polarization. The circularly (linearly) polarized pump pulse, sufficiently strong (typically 1 mW/mm$^2$) and short when compared to Larmor frequency in a given magnetic field, creates atomic polarization (coherence) in the sample. Subsequently, a weak (below 1 μW) probe beam is turned on and its polarization rotation is recorded. The probe may be applied with appropriate delay to enable measurements of the time evolution of the rotation signals for up to some 20 ms (limit imposed by the freefall of atoms). The rotation measurement is followed by switching back the MOT fields. The expanding cloud is then recaptured and the whole sequence is repeated. The typical cycle time lasts a fraction of a second.

### NFE signal buildup, amplitude and width

The characteristics of the NFE resonance strongly depend on the light intensity, as depicted in Fig. S1. Increasing light power causes the resonance to appear larger and faster. The bottom panel illustrates the broadening effect of a resonant probe light at larger intensities. The linear extrapolation of the low-intensity dependence to the zero light power (Fig. S1c) yields about 4.5 mG intercept which corresponds to about 150 μs coherence time. This is significantly shorter than expected from just the free-fall contribution ($\gamma_{TOF} \sim \frac{1}{10\text{ ms}}$), suggesting presence of other decoherence mechanisms. To study them systematically, the light-dependent contribution needs to be minimized. This is not trivial as a certain pump light intensity is necessary for the signal amplitude build-up, particularly in the one beam setup when the probing cannot be separated from pumping.

### Effect of magnetic shielding on FID signals

In Fig. S2 we demonstrate the effect of the magnetic field inhomogeneity on the recorded FID signals of atomic polarization. If the compensation of stray fields is limited to the DC-field components only, the resulting FID curve is quickly damped (within about 200 μs, blue trace). Compensation of the first-order magnetic-field gradients extends the duration of the FID signal about 3 times (red signal). The former and latter FID signals have clearly non-exponential envelopes and/or exhibit some beating which indicate complexity of the relaxation processes. On the other hand, shielding of the science chamber yields regular FID curves (black trace) characterized by lower damping rate of about 1 ms for CW 70 MHz blue detuned probe of 1 μW/mm$^2$ intensity.

### Stroboscopic measurement of the coherence FID

Coherence FID signals have the largest amplitude for probe light close to the resonance (Fig. 4c of the *main paper*). Under such conditions the FID signal decays quickly due to the decoherence caused by interaction with the probe beam. Its influence can be mitigated by replacing continuous wave probing with a pulsed scheme, specifically, by the, so called, *relaxation in the dark* approach[36]. In Figure S3 we present a coherence FID signal recorded by delaying the probe pulse relative the pump one and repeating the measurement over many realizations with several increasing delays. Although individual FID curves (shown with different colors) decay rapidly, the envelope of their maximal amplitudes corresponding to different delays and marked with dashed red lines decays slower. This is due to the fact that within the delay time between pumping and probing, the probe beam does not perturb the atoms (their relaxation occurs in the dark). Consequently, the pulsed (strobed) probing significantly slows down the decoherence, and enables observations of coherence FID signals with the decay times comparable to those seen with population signals.

### Magnetic transients due to the trap switching without the shielding

Instability of both: the magnitude and the orientation of magnetic field may lead to a complex magneto-optical response. Magnetic rotation signals recorded after switching off the MOT quadrupole field and with the unshielded science chamber are visualized in Fig. S4a. The residual magnetic field results from imperfect compensation and inhomogeneity of the laboratory DC field, residual 50 Hz mains noise and decaying eddy-current-induced transients. To slow down the decoherence caused by the probe beam its frequency was blue detuned by 70 MHz. The inset to Fig. S4a presents time-dependent rotation response measured without pumping pulses, i.e. exclusively by the probe beam after switching off the MOT fields at $t = 0$. The black and red traces correspond to triggering the measurement sequence by the mains line with the 0 and 180 deg phase (the rising and falling slopes, respectively). These signals reflect mainly the effect of the longitudinal component, $\theta(B_z) \propto B_z$ of the stray field. The main part of Fig. S4a presents rotation signals acquired with additional resonant, circularly polarized pump pulses repeated every 0.7 ms, for the same triggering conditions as used for the signals in the inset. The rotation signals observed with pump pulses consist of the contributions caused by the longitudinal magnetic-field component $B_z$ of the stray field superimposed with new oscillating contributions following each pump pulse. These contributions are the FID signals due to total magnetic field experienced by atoms at various instants. By measuring the distance between local rotation signal extrema, it is possible to estimate the Larmor precession frequency and, therefore, to read out the full magnitude of the field, $|B|$, even without calibration of the rotation signal (Fig. S4b). The accuracy of such a procedure depends on a number of visible oscillations and a degree of FID signal distortion.

### Residual magnetic instability with the shielding

The single-beam NFE rotation may also be used for diagnostics of small field transients, e.g., associated with switching off the MOT quadrupole field and applying the Faraday field. Figure S5 presents the close up of rotation signal shown in the main text (Fig. 2b), i.e., the contour plot of the rotation signal recorded in the shielded experiment as a function of time and a near-zero longitudinal magnetic field. The asymmetry and drift of rotation signal around zero applied-field reflects transient fields caused by eddy currents arising within the μ-metal shielding, as well as the onset of the applied magnetic field. The initial field instability is smaller than in the unshielded case (Fig. S4), as the probing starts 4 ms after switching off the MOT.

**Table 1**. A reference list for the isotopes that were used for each of the presented measurements.

| Figure | 1a | 2a-d | 3a | 3b | 4a,b | 4c,d | 5a,b | 6a-c | S1-5 |
|---|---|---|---|---|---|---|---|---|---|
| Isotope | $^{85}$Rb | $^{85}$Rb | $^{85}$Rb | $^{87}$Rb | $^{85}$Rb | $^{87}$Rb | $^{87}$Rb | $^{85}$Rb | $^{85}$Rb |

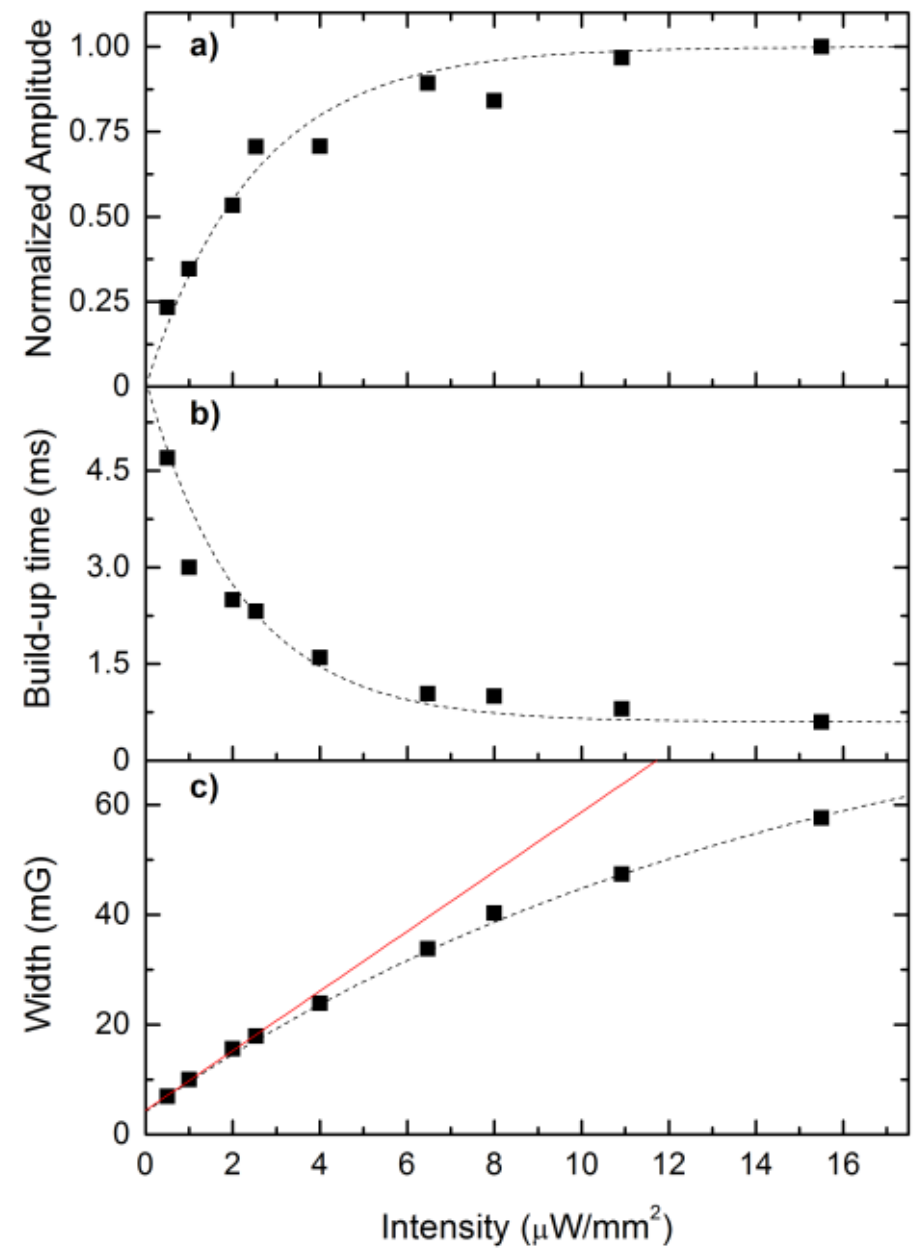

**Fig. S1**. Dependence of (a) amplitude, (b) width and (c) buildup time of the maximum amplitude of the NFE resonance on the resonant light intensity illustrating saturation and power broadening of the rotation resonance.

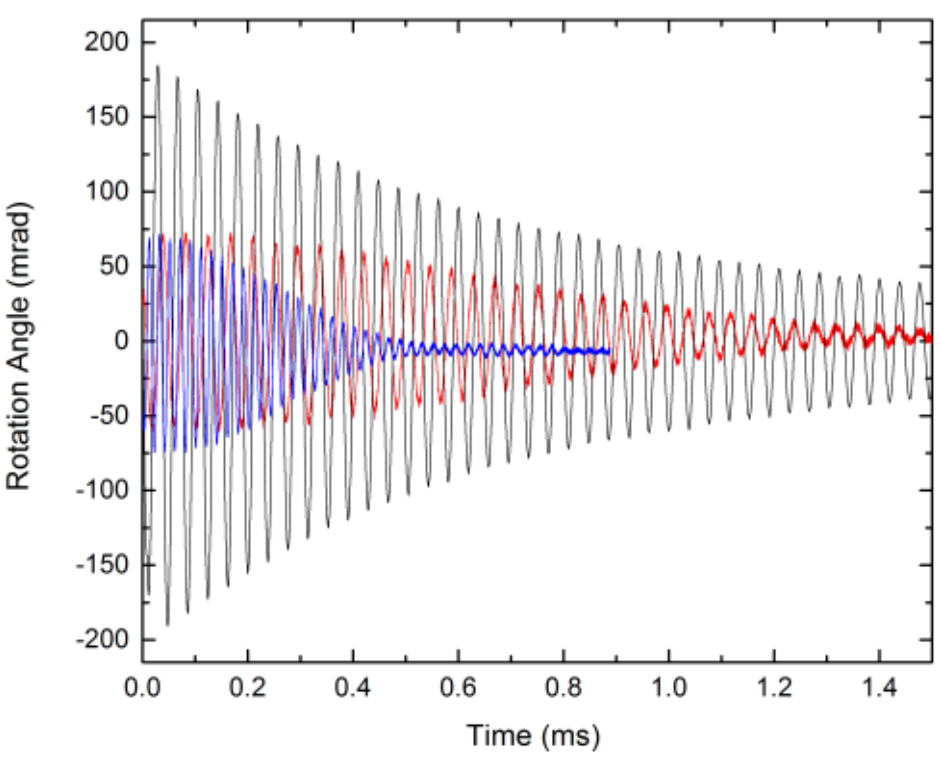

**Fig. S2**. FID signals of atomic polarization $\theta(t)$ recorded by a CW probe with: compensation of DC stray magnetic fields (blue); its DC value and first-order gradients (red); and with no additional compensation but with magnetic shielding of the science cell (black). $B_z = 0$, and $B_x = 56$ mG. Note the change in FID signal shape for the unshielded cases

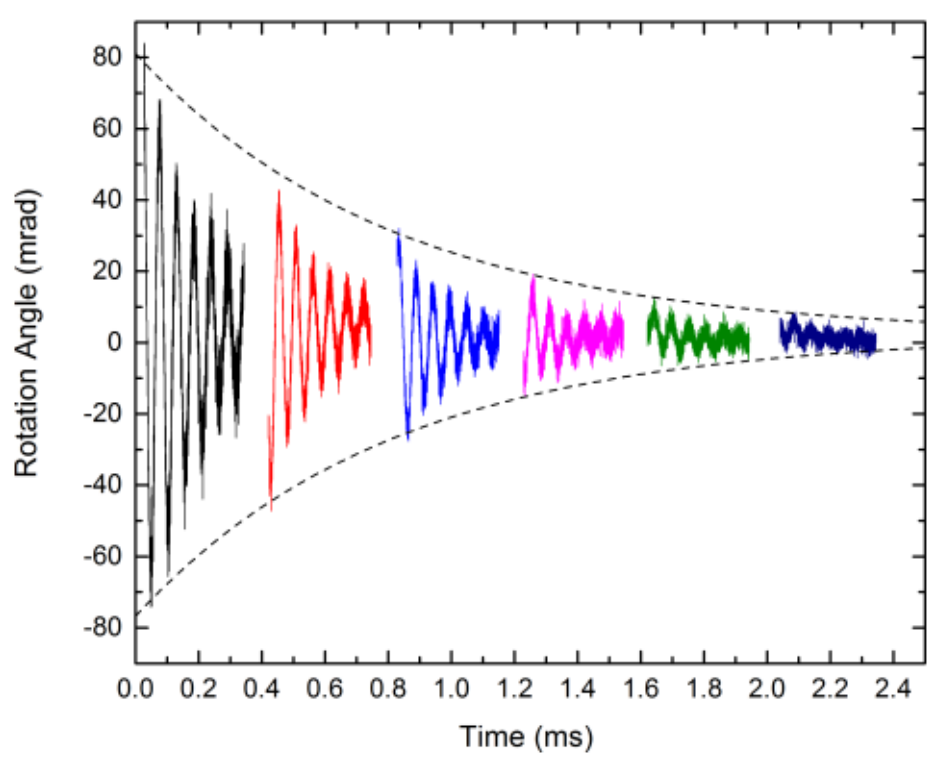

**Fig. S3**. Stroboscopic measurement of the coherence FID. The plot depicts many recordings of the FID oscillations with different delays between the pump and probe pulses (marked with different colors). Although individual FID curves decay rapidly because of the probe-induced decoherence, the envelope of their maximal amplitudes corresponding to different delays decays slower: typical decay time of individual FID oscillations are 170 μs, while that of the envelope of the strobed FID amplitude is 815 μs as shown by dashed lines.

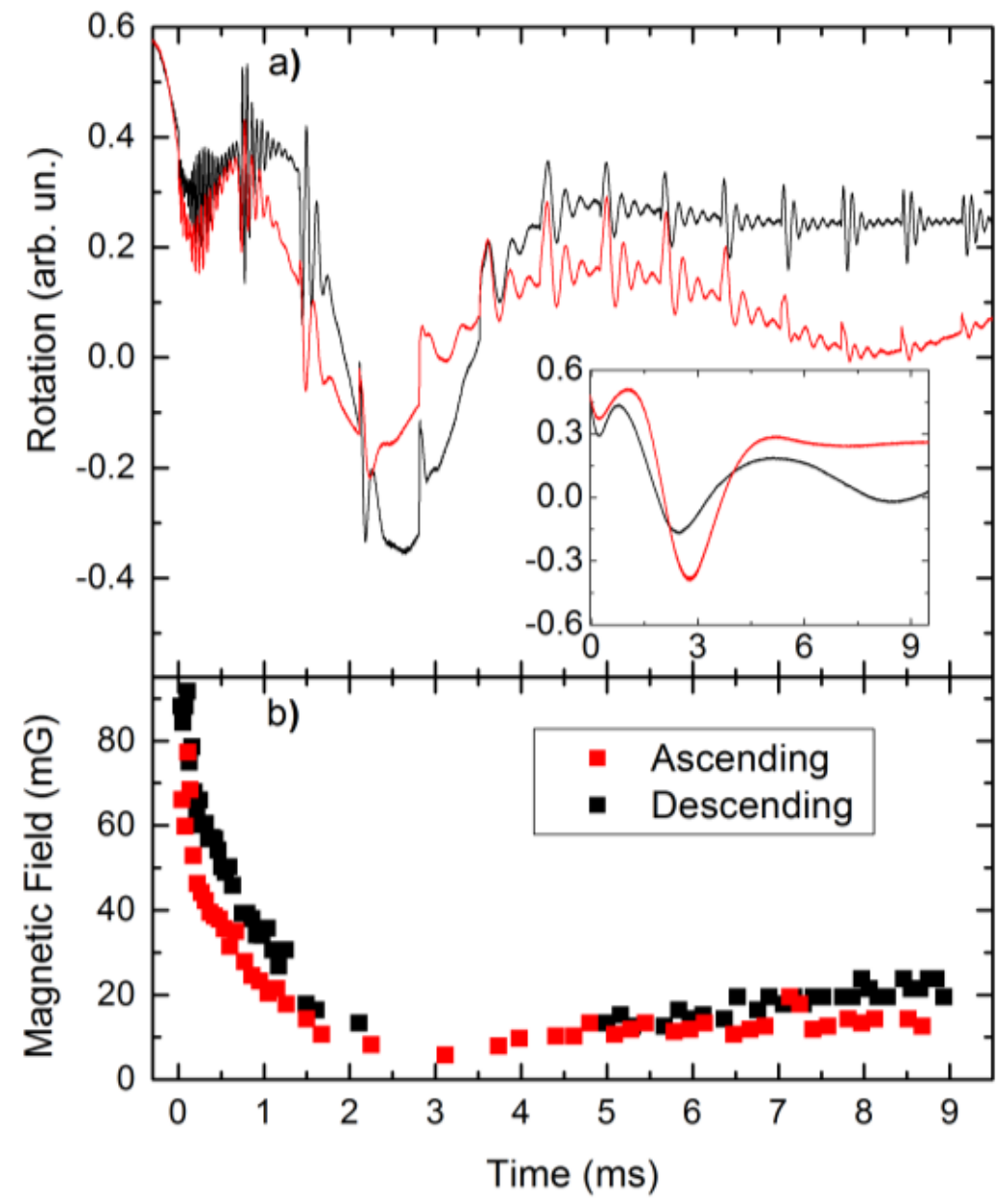

**Fig. S4**. Rotation signal $\boldsymbol{\theta(t)}$ of a CW probe beam recorded after switching off the MOT at $\boldsymbol{t = 0}$ with an unshielded MOT. (a) Two sets of signals are presented: without pumping (inset) and with periodic pumping (main panel). The red and black traces correspond to triggering by the 0 and 180 deg phases of the mains line, respectively. The signal in the inset represents the slowly-varying part in the main frame and reflects the instantaneous value of the longitudinal component of the field vector. (b) The FID oscillations following the pump pulses occur at Larmor frequency and provide a measure of the amplitude $|B|$ of the total magnetic-field. The probe was blue detuned by 70 MHz.

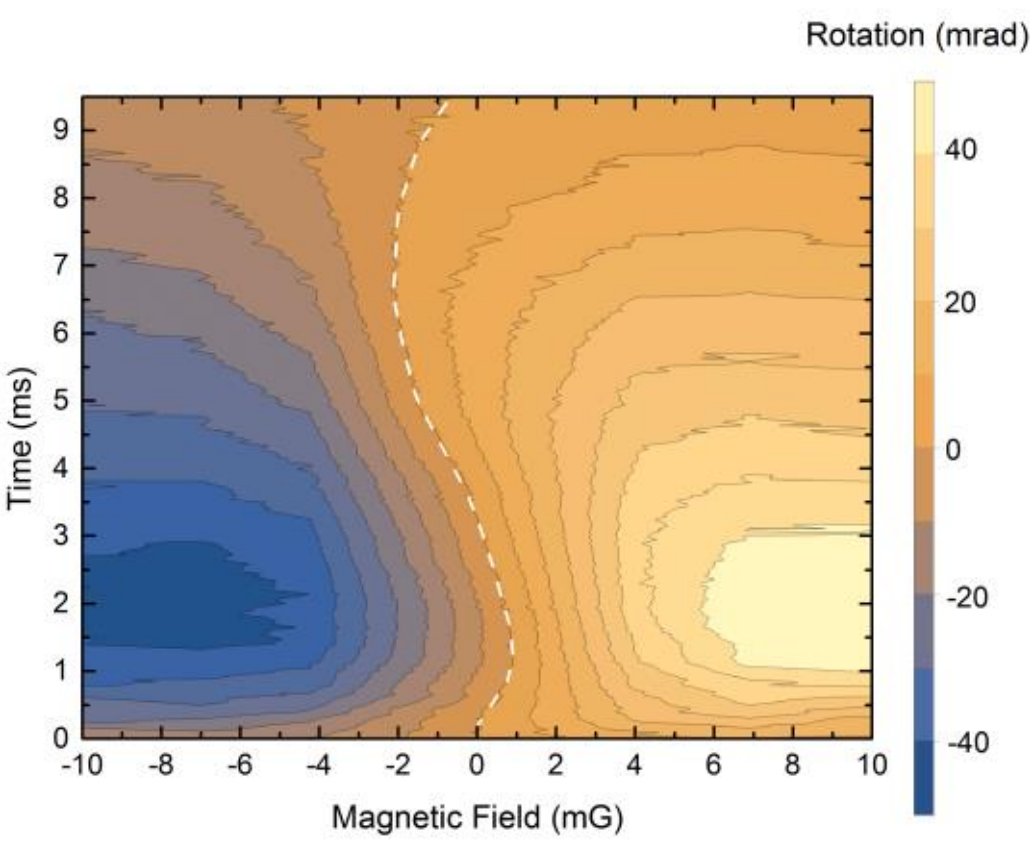

**Fig. S5**. Zoom into rotation signals from Fig. 2b (of the *main paper*) showing in-detail the small drifts of the resonance centre caused by switching from MOT quadrupole field to the uniform Faraday field. White broken line indicates the position of zero rotation value at various times. Signals were recorded with the magnetic shielding.